\documentclass[12pt]{article}
\usepackage{amsmath,a4wide,array,axodraw,multirow,theorem}

\theoremstyle{break}\theorembodyfont{\rmfamily}
                    \newtheorem{Alg}{Algorithm}%[section]
\addtocounter{secnumdepth}{1}

\begin{document}

%% labels %%
\newcommand{\eqn}[1]{Eq.\!~(\ref{#1})}
\newcommand{\fig}[1]{Fig.\!~\ref{#1}}
\newcommand{\tab}[1]{Tab.\!~\ref{#1}}
\newcommand{\Sec}[1]{Section~\ref{#1}}
%
%% greek alphabet %%
\newcommand{\al}{\alpha}
\newcommand{\be}{\beta}
\newcommand{\vt}{\vartheta}
\newcommand{\ep}{\epsilon}
\newcommand{\ve}{\varepsilon}
\newcommand{\vp}{\varepsilon}
\newcommand{\la}{\lambda}
\newcommand{\si}{\sigma}
\newcommand{\de}{\delta}
\newcommand{\vhi}{\varphi}
%
%% sets %%
\newcommand{\Pol}{{\bf P}}
\newcommand{\Real}{{\bf R}}
\newcommand{\Comp}{{\bf C}}
\newcommand{\Inat}{{\bf I}}
%
%% misc %%
\newcommand{\df}{=}
\newcommand{\ra}{\rightarrow}
\newcommand{\lar}{\leftarrow}
\newcommand{\half}{{\textstyle\frac{1}{2}}}
\newcommand{\sfrac}[2]{{\textstyle\frac{#1}{#2}}}
\newcommand{\sgn}{\textrm{sgn}}
\newcommand{\rambo}{\texttt{RAMBO}}
\newcommand{\sarge}{\texttt{SARGE}}
\newcommand{\np}{n}
\newcommand{\wcm}{\sqrt{s}\,}
\newcommand{\scm}{s}
\newcommand{\ip}[2]{#1\!\cdot\!#2}
\newcommand{\ipb}[2]{(#1#2)}
\newcommand{\inip}[2]{\{#1|#2\}_\de}
\newcommand{\Bo}{\mathcal{H}}
\newcommand{\Ro}{\mathcal{R}}
\newcommand{\Lo}{\mathcal{L}}
\newcommand{\Aqcd}{A^{\textrm{QCD}}}
\newcommand{\Ant}{A^{\scriptscriptstyle\textrm{QCD}}_{\scm}}
\newcommand{\pin}{0}
\newcommand{\nip}{\tilde{0}}
\newcommand{\ppin}{p_0}
\newcommand{\pnip}{\tilde{p}_{0}}
\newcommand{\xim}{\xi_{\textrm{m}}}
\newcommand{\xilow}{\xi_{\textrm{low}}}
\newcommand{\enul}{e_0}
\newcommand{\ethr}{e_3}
\newcommand{\invs}[2]{{#1}^{2}_{#2}}
\newcommand{\invm}[1]{m_{#1}}
\newcommand{\tcpu}{t_{\textrm{cpu}}}
\newcommand{\pvec}{\hspace{-1pt}\vec{\hspace{1pt}p}}
\newcommand{\qvec}{\hspace{-1pt}\vec{\hspace{1pt}q}}
\newcommand{\dfp}{d^4\hspace{-1pt}p}
\newcommand{\dfq}{d^4\hspace{-1pt}q}
\newcommand{\dfk}{d^4\hspace{-1pt}k}
\newcommand{\gl}{\textrm{g}}
\newcommand{\GeV}{\textrm{GeV}}
\newcommand{\Nge}{N_{\textrm{ge}}}
\newcommand{\Nac}{N_{\textrm{ac}}}
\newcommand{\ee}[1]{\!\times\!10^{#1}}
\newcommand{\emu}[1]{\!\times\!10^{-#1}}
\newcommand{\hour}{\textrm{h}}
\newcommand{\seco}{\textrm{s}}
\newcommand{\texa}{t_{\textrm{exa}}}
\newcommand{\Cap}[1]{\caption{\footnotesize #1}}

\title{{\bf Generating QCD-antennas}}

\author{Andr\'e van Hameren\thanks{{\tt andrevh@sci.kun.nl}, 
                                        $^\dagger${\tt kleiss@sci.kun.nl}} ~and 
        Ronald Kleiss$^\dagger$\\
        University of Nijmegen, Nijmegen, the Netherlands}

\maketitle

\begin{abstract}
An extension of the \sarge-algorithm of \cite{HKD} is introduced, which
includes the incoming momenta in the kinematical pole structure of the density
with which the momenta are generated. The algorithm is compared with \rambo\
in the integration of QCD-amplitudes in the SPHEL-approximation, and the
computing times are extrapolated to those for the calculation with exact matrix
elements.
\end{abstract}

%\thispagestyle{empty}

%\newpage
\section{Introduction}
In future experiments with hadron colliders, such as the LHC, many multi-jet
events will occur.  
%
% The generic situation is depicted schematically in \fig{SarFig01}. Time can
% be considered to flow from the left to the right in the picture. The hadrons
% start the interaction with the emission of partons. The transition of a
% hadron into the emitted parton and the leftover is represented by the white
% blobs. This is the first step. In the second step, represented by the grey
% blob, the partons interact, resulting in $\np$ new partons. In step three,
% these partons turn into jets with high particle multiplicities.
% \begin{figure}[t]
% \begin{center}
% \begin{picture}(160,140)(0,13)
% %
% \ArrowLine(-20,135)(20,120)\Text(-15,125)[r]{hadron}
% \ArrowLine(20,120)(80,80)
% \ArrowLine(20,120)(50  ,150)
% \ArrowLine(20,120)(59.2,136.2)
% %\ArrowLine(20,120)(62.4,120)
% %
% \ArrowLine(20,40)(80,80)
% \ArrowLine(20,40)(50  ,10)
% \ArrowLine(20,40)(59.2,23.8)
% %\ArrowLine(20,40)(62.4,40)
% \ArrowLine(-20,25)(20,40)\Text(-15,35)[r]{hadron}
% %
% \ArrowLine(80,80)(140,140)\Text(145,145)[l]{jet $1$}
% \ArrowLine(80,80)(158.4,112.5)\Text(163.4,118)[l]{jet $2$}
% \DashCArc(80,80)(55,318.5,17){3}
% \ArrowLine(80,80)(140,20)\Text(145,17)[l]{jet $\np$}
% %
% \GCirc(80,80){20}{0.7}
% \BCirc(20,120){10}
% \BCirc(20,40){10}
% \end{picture}
% \Cap{A generic multi-jet event}
% \label{SarFig01}
% \end{center}
% \end{figure}
% The idea is that the contribution of the three steps more-or-less factorize 
% in the transition amplitude of the whole event, and that the processes can 
% be dealt with separately. In this chapter, we deal 
% with step two, the grey blob. 
% 
These can be divided into interesting events (IE), and the background. The main
difference between the two classes is that the Standard Model shall not have
proven yet its capability of dealing with the description of the IE at the
moment when they are analyzed. The background shall not manifest itself as
such a heavy test for the standard model. However, we still need to know the
cross sections of the background in order to compare the ratio of these and
those of the IE with the predictions of the Standard Model. 

For large part of the background, a piece of the transition amplitude consists
of a multi-parton QCD-amplitude, and it is well known \cite{Kuijf} that it
contributes to the cross section with a singular behavior in phase space (PS),
given by the so-called {\em antenna pole structure} (APS). In particular, for
processes involving only $n$ gluons the most important contribution comes from
the sum of all permutations in the momenta of 
\begin{equation}
   \frac{1}{(\ip{p_1}{p_2})(\ip{p_2}{p_3})(\ip{p_3}{p_4})\cdots
            (\ip{p_{\np-1}}{p_{\np}})(\ip{p_{\np}}{p_1})} \;\;,
\label{SarEq001}	    
\end{equation}
and the singular nature stems from the fact that the scalar products
$\ip{p_i}{p_j}$ can become very small.  For the calculation of integrals over
PS, the Monte Carlo (MC) method is the only option, so that an algorithm to
generate random momenta is needed. For processes at high energy, the momenta
may be massless, and \rambo\ \cite{SKE} generates any number of them
distributed uniformly in PS.  This uniform distribution, however, has the
disadvantage that, for the integration of an integrand containing the APS, a
large number of events is needed to reach a result to acceptable precision.  As
an illustration, we present in the left table of \tab{SarTab01} the number of
PS points needed to integrate the single antenna of \eqn{SarEq001}, so not even
the sum of its permutations, to an expected error of $1\%$. 
\begin{table}[t]
\begin{center}
\begin{tabular}{|c|c|c|}
  \hline
  \multicolumn{3}{|c|}{single antenna integrated to $1\%$ error}\\ \hline\hline
   \multirow{2}{19mm}{number of momenta}
  &\multirow{2}{21mm}{$\displaystyle\frac{\textrm{cut-off}}
                                         {\textrm{CM-energy}}$} 
  &\multirow{2}{19mm}{number of PS points} \\ 
  & & \\ \hline
  $3$ & $0.183$ & $10,069$ \\ \hline
  $4$ & $0.129$ & $26,401$ \\ \hline
  $5$ & $0.100$ & $58,799$ \\ \hline
  $6$ & $0.0816$ & $130,591$ \\ \hline
  $7$ & $0.0690$ & $240,436$ \\ \hline
  $8$ & $0.0598$ & $610,570$ \\ \hline
\end{tabular}
\qquad
\begin{tabular}{|c|c|c|}
  \hline
  \multicolumn{3}{|c|}{evaluation amplitude in $1$ PS point}\\ \hline\hline
   \multirow{2}{21mm}{number of final gluons}
  &\multicolumn{2}{c|}{cpu-time (seconds)}\\ \cline{2-3}
  & {\tt SPHEL} & exact \\ \hline
%  $2$ & $1.26\emu{5}$ & $4.09\emu{2}$\\ \hline
  $3$ & $2.83\emu{5}$ & $1.60\emu{1}$ \\ \hline
  $4$ & $9.76\emu{5}$ & $5.54\emu{1}$ \\ \hline
  $5$ & $4.88\emu{4}$ & $1.945$ \\ \hline
  $6$ & $3.26\emu{3}$ & $6.06$ \\ \hline
  $7$ & $2.57\emu{2}$ & $19.91$ \\ \hline
  $8$ &                    & $64.45$ \\ \hline
\end{tabular}
\Cap{Typical number of PS points and computing times.}
\label{SarTab01}
\end{center}
\vspace{-20pt}
\end{table}
The antenna cannot be integrated over the whole of PS because of the
singularities, so these have to be cut out. This is done through the restriction
$\invs{(p_i+p_j)}{}\geq s_0$ for all $i,j=1,\ldots,\np$,
% \footnote{Remember that $\invs{(p+q)}{}=2\ip{p}{q}$ since $p$ and $q$ are 
% massless.} 
and in the table 
the ratio between $\sqrt{s_0}$ and the total energy $\wcm$ is given. These
numbers are based on the reasonable choice $s_0/s=0.2/[n(n-1)]$.

Performing MC integration with very many events is not a problem if the
evaluation of the integrand in each PS point is cheap in computing time.  This
is, for example, the case for algorithms to calculate the squared multi-parton
amplitudes based on the so called SPHEL-approximation, for which only the
kinematical structure of (\ref{SarEq001}) is implemented \cite{Kuijf}.
Nowadays, algorithms to calculate the {\em exact} matrix elements exist, which
are far more time-consuming \cite{DKP,CMMP}. As an illustration of what is
meant by `more time-consuming', we present the right table of \tab{SarTab01}
with the typical cpu-time needed for the evaluation in {\em one} PS point of
the integrand for processes of two gluons going to more gluons, both for the
SPHEL-approximation and the exact matrix elements \cite{Petros}. It is
expected, and observed, that the exact matrix elements reveal the same kind of
singularity structures as the APS, so that, according to the tables, the PS
integration for a process with $8$ final gluons would take in the order of $400$
days\ldots

The solution to this problem is importance sampling. Instead of \rambo, a PS
generator should be used which generates momenta with a density including the
APS. In \cite{HKD}, we introduced an algorithm that does part of the job, and
is called \sarge\ (from {\tt S}taggered {\tt A}ntenna {\tt R}adiation
{\tt G}enerator). It generates $n$ random momenta with a density proportional 
to the the sum of all permutations of (\ref{SarEq001}), and because they are 
all random, they should be interpreted as outgoing momenta. In the pole 
structure of a ``real'' QCD-amplitude, however, also the incoming momenta 
occur. In this paper, we introduce an extension of the \sarge-algorithm, which
includes these pole structures. We compare it with \rambo\ in the 
calculation of integrals of QCD-amplitudes in the SPHEL-approximation, and 
extrapolate the computing times to those for the calculation with exact matrix 
elements. The conclusion will be that \sarge\ takes account for a
substantial reduction in computing time. 

For the sake of completeness, we describe the full algorithm in this paper,
including the piece that was introduced in \cite{HKD}. What we actually
presented there was only the algorithm and no proof of its correctness
whatsoever. In this paper, however, we shall meet our engagements with respect
to this. We do this with the help of the unitary algorithm formalism, which we
introduce in the following section by its application to the \rambo-algorithm.

\section{Notation and the unitary algorithm formalism}
The relativistic momentum $p = (p^0,p^1,p^2,p^3) \df (p^0,\pvec)$
of an elementary particle is a vector in $\Real^4$.
The momentum with the opposite $3$-momentum is denoted by
\begin{equation}
   \tilde{p}\df(p^0,-\pvec)  \;\;.
\end{equation}
We shall need the first and the fourth canonical basis vectors, which we denote 
\begin{equation}
   \enul\df(1,0,0,0) \qquad\textrm{and}\qquad \ethr\df(0,0,0,1)  \;\;.
\end{equation}
A typical parameterization of a $3$-momentum with unit length is given by 
$\hat{n}(z,\vhi)$, where
\begin{equation}
   \hat{n}_1(z,\vhi) \df \sqrt{1-z^2}\sin\vhi\;\;,\quad
   \hat{n}_2(z,\vhi) \df \sqrt{1-z^2}\cos\vhi\;\;,\quad
   \hat{n}_3(z,\vhi) \df z \;\;.
\label{DefHatn}   
\end{equation}
The Lorentz invariant scalar product shall be denoted with a dot or with 
parentheses:
\begin{equation}
   \ipb{p}{q} \df
   \ip{p}{q} \df p^0q^0 - \ip{\pvec}{\qvec} \quad,\qquad
   \ip{\pvec}{\qvec} \df p^1q^1 + p^2q^2 + p^3q^3 \;.
\end{equation}
The product of a vector with itself is denoted as a square
\begin{equation}
   \invs{p}{} \df \ipb{p}{p} = (p^0)^2 - |\pvec|^2  \quad,\qquad
   |\pvec|\df(\ip{\pvec}{\pvec})^{1/2} \;\;.
\end{equation}
The same notation for the quadratic form and the $2$-component will not lead 
to confusion, because the $2$-component will not appear explicitly anymore
after this section.
For physical particles, $\invs{p}{}$ has to be positive, and in that case, the
square root gives the invariant mass of the particle:
\begin{equation}
   \invm{p} \df \sqrt{\invs{p}{}} \quad\textrm{if}\;\; \invs{p}{}\geq0 \;\;.
\end{equation}
A boost that
transforms a momentum $p$, with $\invs{p}{}>0$, to $\invm{p}\enul$ is
denoted $\Bo_p$, so 
\begin{equation}
   \Bo_pp = \invm{p}\enul \qquad\textrm{and}\qquad 
   \invm{p}\Bo_p\enul = \tilde{p}  \;\;.
\end{equation}
% More explicitly, such a boost is given by
% \begin{equation}
%    \Bo_pq
%    = (a,\qvec-b\pvec)
%    \qquad\textrm{where}\qquad
%    a = \frac{\ip{p}{q}}{\invm{p}} \;\;,\quad 
%    b = \frac{q^0+a}{p^0+\invm{p}}\;\;.
% \end{equation}
A rotation that transforms $p$ to $p^0\enul+|\pvec|\ethr$ is denoted $\Ro_p$,
so 
\begin{equation}
   \Ro_pp = p^0\enul+|\pvec|\ethr \qquad\textrm{and}\qquad
   \Ro_p\tilde{p} = p^0\enul-|\pvec|\ethr\;\;.
\end{equation}
Since rotations only change the $3$-momentum, we shall use the same symbol if 
a rotation is restricted to three-dimensional space.

The physical PS of $\np$ particles is the $(3\np-4)$-dimensional
subspace of $\Real^{4\np}$, given by the restrictions that the energies of the 
particles are positive, the invariant masses squared
$\invs{p}{i}$ are fixed to given positive values $s_i$, and that
the sum 
\begin{equation}
   p_{(\np)}\df\sum_{i=1}^{\np}p_i 
\end{equation}
of the momenta is fixed to a given momentum $P$. The restrictions for the
separate momenta shall be expressed with a `PS characteristic distribution'
\begin{equation}
   \vt_{s_i}(p) \df \delta(\invs{p}{}-s_i)\,\theta(p^0)\;\;, 
   \qquad\textrm{and}\qquad
   \vt(p) \df \vt_0(p) \;\;.
\end{equation}
The generic PS
integral, of a function $F$ of a set $\{p\}_\np\df\{p_1,\ldots,p_\np\}$ of
momenta, that has to be calculated is then given by 
\begin{equation}
   \int_{\Real^{4\np}}
   \Big(\prod_{i=1}^{\np}\dfp_i\,\vt_{s_i}(p)\Big)\, 
   \de^4(p_{(\np)}-P)\,F(\{p\}_\np) \;\;.
\end{equation}
An integral shall always start with a single $\int$-symbol, and for every 
integration variable, say $z$, a $dz$ means `integrate $z$ over the 
appropriate integration region'. If it is not evident what this region is, it 
shall be made explicit with the help the of logical $\theta$-functions, 
which have statements $\Pi$ as argument, and are defined through
\begin{equation}
   \theta(\Pi) \df \begin{cases}
                   1 &\textrm{if $\Pi$ is true}\\
		   0 &\textrm{if $\Pi$ is false.}
		   \end{cases}
\end{equation}

\subsection{The \rambo\ algorithm in the UAF}
\rambo\ was developed with the aim to generate the flat PS distribution of
$\np$ massless momenta as uniformly as possible, and such that the sum of the
momenta is equal to $\wcm\enul$ with $\scm$ a given squared energy. This
means that the system of momenta is in its center-of-mass frame (CMF), and that
the density is proportional to the `PS characteristic distribution' 
\begin{equation}
    \Theta_{\scm}(\{p\}_\np) \;\df\; 
   \delta^4(p_{(\np)}-\wcm\enul)\prod_{i=1}^{\np}\vt(p_i) \;\;.
\label{RamEq002}   
\end{equation}
The algorithm consists of the following
steps:
\begin{Alg}[\texttt{RAMBO}]
\begin{enumerate}
\item generate $\np$ massless vectors $q_j$ with positive energy without 
      constraints but under some normalized density $f(q_j)$;
\item compute the sum $q_{(\np)}$ of the momenta $q_j$;
\item determine the Lorentz boost and scaling transform that
      bring $q_{(\np)}$ to $\wcm\enul$;
\item perform these transformations on the $q_j$, and call the result $p_j$.
\end{enumerate}
\label{AlgRambo}
\end{Alg}
Trivially, the algorithm generates momenta that satisfy the various
$\de$-constraints, but it is not clear a priori that the momenta have the
correct distribution. To prove that they actually do, we apply the unitary
algorithm formalism (UAF). We write the generation of a variable as the
integral of the density with which that variable is generated, and interpret
every assignment as a generation with a density that is given by a Dirac
delta-distribution. Only the assignment of the final output should not be
written as an integral, but only with the delta-distributions. The UAF tells
us that Algorithm \ref{AlgRambo} generates a density 
\begin{align}
%   \Phi_{\scm}(\{p\}_\np) = 
%   \int &\prod_{j=1}^{\np}\dfq_j\vt(q_i)f(q_j)
%         \notag\\
%        &\times d^4b\,\de^4(b-q_{(\np)}/\invm{q_{(\np)}})\,
%          dx\,\de(x-\wcm/\invm{q_{(\np)}})\, \notag\\
%        &\times \prod_{i=1}^{\np}\de^4(p_i-x\Bo_bq_i) \;\;.
   \Phi_{\scm}(\{p\}_\np) = 
   \int &\Big(\prod_{j=1}^{\np}\dfq_j\vt(q_i)f(q_j)\Big)\,
         d^4b\,\de^4\Big(b-\frac{q_{(\np)}}{\invm{q_{(\np)}}}\Big)\,
          dx\,\de\Big(x-\frac{\wcm}{\invm{q_{(\np)}}}\Big)\, \notag\\
        &\times \prod_{i=1}^{\np}\de^4(p_i-x\Bo_bq_i) \;\;.
\label{RamEq001}
\end{align}
The unitarity of the algorithm is expressed by the fact that integration of the
above equation over the set of variables $\{p\}_\np$ leads to the identity
$1=1$.
To calculate the distribution yielded by this algorithm, the integral has to be
evaluated. First of all, some simple algebra using
$p_{(\np)}=x\Bo_bq_{(\np)}$, $q_{(\np)}=x^{-1}\Bo_b^{-1}p_{(\np)}$ and the
Lorentz and scaling properties of the Dirac $\de$-distributions leads to 
\begin{equation}
   \de^4\Big(b-\frac{q_{(\np)}}{\invm{q_{(\np)}}}\Big)\,
   \de\Big(x-\frac{\wcm}{\invm{q_{(\np)}}}\Big)
   \;=\; \frac{2\scm^2}{x}\,\de^4(p_{(\np)}-\wcm\enul)\,\de(\invs{b}{}-1) \;\;.
\end{equation}
Furthermore, since we may write
\begin{equation}
  \dfq_j\,\de(\invs{q}{j})\,\de^4(p_j-x\Bo_bq_j) 
  \;=\; \frac{1}{x^2}\,\de(\invs{p}{j}) 
\end{equation}
under the integral, the l.h.s. of \eqn{RamEq001} becomes
\begin{equation}
   \int\Theta_{\scm}(\{p\}_\np)\,d^4b\,\de(\invs{b}{}-1)\,
           dx\frac{2\scm^2}{x^{2\np+1}}
	              \prod_{i=1}^{\np}f(\frac{1}{x}\Bo_b^{-1}p_i)\,
           \theta(\ip{\enul}{\Bo_b^{-1}p_j}>0) \;.
\end{equation}
In the standard \rambo\ algorithm, the following choice is made for $f$:
\begin{equation}
  f(q) \;=\; \frac{c^2}{2\pi}\,\exp(-cq^0) \;\;,
\end{equation}
where $c$ is a positive number with the dimension of an inverse mass.
Therefore, if we use that $p_{(\np)}=\wcm\enul$ and that $q^0=\ip{\enul}{q}$
for any $q$, then 
\begin{align}
   \prod_{i=1}^{\np}f(\frac{1}{x}\Bo_b^{-1}p_i)\,
                     \theta(\ip{\enul}{\Bo_b^{-1}p_i}>0)
   \;=\; &\left(\frac{c^2}{2\pi}\right)^{\np}
          \exp\left(-\frac{c}{x}\,\ip{\enul}{\Bo_b^{-1}p_{(\np)}}\right) \,
          \prod_{i=1}^{\np}\theta(\ip{\enul}{\Bo_b^{-1}p_i}>0)  \notag\\
   \;=\; &\left(\frac{c^2}{2\pi}\right)^{\np}\,
          \exp\left(-\frac{c\wcm}{x}\,b^0\right)\,\theta(b^0>0) \;\;. 
\end{align}
As a result of this, the variables $p_i$, $i=1,\ldots,\np$ only appear in
$\Theta_{\scm}$, as required.  The remaining integral is calculated in Appendix
A, with the result that \rambo\ generates the density 
\begin{equation}
   \Phi_{\scm}(\{p\}_\np) 
   \;=\; \Theta_{\scm}(\{p\}_\np)\left(\frac{2}{\pi}\right)^{\np-1}
         \frac{\Gamma(\np)\Gamma(\np-1)}{\scm^{\np-2}} \;\;.		      
\end{equation}
Incidentally, we have computed here the volume of the
PS for $n$ massless particles:
\begin{equation}
   \int_{\Real^{4\np}} d^{4n}\hspace{-1pt}p\,\Theta_{\scm}(\{p\}_\np) 
   \;=\; \left(\frac{\pi}{2}\right)^{\np-1}
                   \frac{\scm^{\np-2}}{\Gamma(\np)\Gamma(\np-1)}\;\;.
\end{equation}
Note, moreover, that $c$ does not appear in the final answer; this is
only natural since any change in $c$ will automatically be compensated
by a change in the computed value for $x$. Finally, it is important
to realize that the `original' PS has dimension $3\np$, while
the resulting one has dimension $3\np-4$: there are configurations
of the momenta $q_j$ that are different, but after boosting and
scaling end up as the same configuration of the $p_j$. It is this
reduction of the dimensionality that necessitates the integrals over
$b$ and $x$.

% The first step of the algorithm consists of generating massless momenta with 
% positive energy. To generate such momenta, we use that 
% \begin{align}
%    \dfp\vt(p)
%    \;=\; d\vhi\,dz\,dp^0p^0\,
%        \theta(p^0>0)\,\theta(0\leq\vhi\leq2\pi)\,\theta(-1\leq z\leq1) \;\;,
% \end{align}
% % with $p^1=p^0\sqrt{1-z^2}\sin\vhi$, $p^2=p^0\sqrt{1-z^2}\cos\vhi$ and
% % $p^3=p^0z$. 
% with $p = (\,p^0,p^0\hat{n}(z,\vhi)\,)$, where 
% \begin{equation}
%    \hat{n}_1(z,\vhi) \df \sqrt{1-z^2}\sin\vhi\;\;,\quad
%    \hat{n}_2(z,\vhi) \df \sqrt{1-z^2}\cos\vhi\;\;,\quad
%    \hat{n}_3(z,\vhi) \df z \;\;.
% \label{DefHatn}   
% \end{equation}
% From this we can directly see that, to generate $p$ following a
% density proportional to $\vt(p)f(p^0)$, one should
% \begin{Alg}[\texttt{MASSLESS MOMENTUM}]
% \begin{enumerate}
% \item generate $p^0$ in $[0,\infty)$ following a density $\sim p^0f(p^0)$;
% \item generate $\vhi$ uniformly distributed in $[0,2\pi]$ and $z$ uniformly 
%       in $[-1,1]$;
% \item construct $\hat{n}(z,\vhi)$ and put $\pvec\lar p^0\hat{n}(z,\vhi)$.
% \end{enumerate}
% \end{Alg}
% To generate $p^0$ following the density $p^0\exp(-p^0)$, one can
% \begin{Alg}[\texttt{0-COMPONENT}]
% \begin{enumerate}
% \item generate $x_1$ and $x_2$ distributed uniformly in $[0,1]$;
% \item put $p^0\lar -\log(x_1x_2)$, 
% \end{enumerate}
% \end{Alg}
% since
% \begin{equation}
%    \int dx_1dx_2\,\theta(0\leq x_{1,2}\leq1)\,\de(\,p^0+\log(x_1x_2)\,)
%    \;=\; \int_{e^{-p^0}}^1 dx\,\frac{e^{-p^0}}{x}
%    \;=\; p^0e^{-p^0} \;.
% \end{equation}

\section{The basic antenna}
As mentioned before, we want to generate momenta that represent radiated
partons with a density that has the antenna structure
$[\ipb{p_1}{p_2}\ipb{p_2}{p_3}\ipb{p_3}{p_4}\cdots,
  \ipb{p_{\np-1}}{p_{\np}}\ipb{p_{\np}}{p_1}]^{-1}$.
Naturally, the momenta can be viewed as coming from a splitting process: one
starts with two momenta, a third is radiated off creating a new pair of momenta
of which a fourth is radiated off and so on. In fact, models similar to this
are used in full-fledged Monte-Carlo generators like {\tt HERWIG}. Let us
therefore first try to generate a single massless momentum $k$, radiated from a
pair of given massless momenta $p_1$ and $p_2$. In order for the distribution
to have the correct infrared and collinear behavior, it should qualitatively be
proportional to $[\ipb{p_1}{k}\ipb{k}{p_2}]^{-1}$. Furthermore, we want the
density to be invariant under Lorentz transformations and scaling of the
momenta, keeping in mind that the momenta are three out of possibly more in a
CMF and that we have to perform these transformations in the end, like in
\rambo.
This motivates us to define the {\em basic antenna\/} structure as
\begin{equation}
   dA(p_1,p_2;k) 
   \;\df\; \dfk\vt(k)\,
           \frac{1}{\pi}\,\frac{\ipb{p_1}{p_2}}{\ipb{p_1}{k}\ipb{k}{p_2}}\;
           g\left(\frac{\ipb{p_1}{k}}{\ipb{p_1}{p_2}}\right)
           g\left(\frac{\ipb{k}{p_2}}{\ipb{p_1}{p_2}}\right)\;\;.
\label{SarEq002}
\end{equation}
Here, $g$ is a function that serves to regularize the infrared and collinear
singularities, as well as to ensure normalization over the whole space for $k$:
therefore, $g(\xi)$ has to vanish sufficiently fast for both $\xi\to0$ and
$\xi\to\infty$. To find out how $k$ could be generated, we evaluate $\int dA$
in the CMF of $p_1$ and $p_2$. Writing 
\begin{equation}
   E\df\sqrt{\ipb{p_1}{p_2}/2} \quad,\qquad
   p\df\Bo_{p_1+p_2}p_1     \quad,\qquad
   q\df\Bo_{p_1+p_2}k \;\;,
\end{equation}
we have
\begin{equation}
   \ipb{p_1}{p_2} = 2E^2              \;\;,\quad
   \ipb{p_1}{k} = Eq^0(1-z)  \;\;,\quad
   \ipb{k}{p_2} = Eq^0(1+z)  \;\;,
\end{equation}
where $z=\ip{\pvec}{\qvec}/(|\pvec||\qvec|)$. The azimuthal angle
of $\qvec$ is denoted $\vhi$, so that
$\qvec=|\qvec|\Ro_{p}^{-1}\hat{n}(z,\vhi)$.  
We can write 
\begin{equation}
    \dfk\vt(k) \;=\; \sfrac{1}{2}\,q^0dq^0\,d\vhi\,dz
               \;=\; \sfrac{1}{2}\ipb{p_1}{p_2}\,d\vhi\,d\xi_1d\xi_2\;\;,
\end{equation}
where,  
\begin{equation}
   \xi_1 = \frac{\ipb{p_1}{k}}{\ipb{p_1}{p_2}} \quad\quad\textrm{and}\quad\quad
   \xi_2 = \frac{\ipb{k}{p_2}}{\ipb{p_1}{p_2}} \;\;,
\end{equation} 
so that $z=(\xi_2-\xi_1)/(\xi_2+\xi_1)$ and $q^0=E(\xi_2+\xi_1)$.
The integral over $dA$ takes on the particularly simple form
\begin{equation}
   \int dA(p_1,p_2;k)
   \;=\; \left(\int_0^\infty d\xi\,\frac{1}{\xi}\,g(\xi)\right)^2\;\;.
\end{equation}
The antenna $dA(p_1,p_2;k)$ will therefore correspond to a unitary algorithm
when we let the density $g$ be normalized by
\begin{equation}
  \int_0^\infty d\xi\,\frac{1}{\xi}\,g(\xi) = 1\;\;.
\end{equation}
Note that the normalization of $dA$ fixes the overall factor uniquely: in
particular the appearance of the numerator $\ipb{p_1}{p_2}$ is forced upon us
by the unitarity requirement.

For $g$ we want to take, at this point, the simplest possible function we can
think of, that has a sufficiently regularizing behavior. We introduce a
positive non-zero number $\xim$ and take
\begin{equation}
   g(\xi) \;\df\; \frac{1}{2\log\xim}\,\theta(\xim^{-1}\leq\xi\leq\xim) \;\;.
\label{SarEq006}   
\end{equation}
The number $\xim$ gives a cut-off for the quotients $\xi_1$ and $\xi_2$ of the
scalar products of the momenta, and not for the scalar products themselves.  It
is, however, possible to relate $\xim$ to the total energy $\wcm$ in the
CMF and a cut-off $s_0$ on the invariant masses, i.e., the requirement that 
\begin{equation}
   \invs{(p_i+p_j)}{} \geq s_0 \qquad \textrm{for all momenta $p_i\neq p_j$.}
   \label{SarEq014}
\end{equation}
This can be done by choosing 
\begin{equation}
   \xim \;\df\; \frac{\scm}{s_0} - \frac{(\np+1)(\np-2)}{2} \;\;.
\label{SarEq007}   
\end{equation}
With this choice, the invariant masses $\invs{(p_1+k)}{}$ and
$\invs{(k+p_2)}{}$ are regularized, but can still be smaller than $s_0$ so that
the whole of PS, cut by (\ref{SarEq014}), is covered. The $s_0$
can be derived from physical cuts $p_T$ on the transverse momenta and
$\theta_0$ on the angles between the outgoing momenta:
\begin{equation}
   s_0 \;=\;
   2p_T^2\cdot\min\left(1-\cos\theta_0\,,\,
                        \left(1+\sqrt{1-p_T^2/\scm}\right)^{-1}\right) \;\;.
\label{SarEq018}			
\end{equation}
With this choice, PS with the physical cuts is covered by PS with the cut of
(\ref{SarEq014}). To generate the physical PS, the method of hit-and-miss Monte
Carlo can be used, i.e, if momenta of an event do not satisfy the cuts, the
whole event is rejected.  We end this section with the piece of the PS
algorithm that corresponds to the basic $dA(p_1,p_2;k)$:
\begin{Alg}[\texttt{BASIC ANTENNA}]
\begin{enumerate}
\item given $\{p_1,p_2\}$, put $p\lar\Bo_{p_1+p_2}p_1$ 
      and put $E\lar\sqrt{\ipb{p_1}{p_2}/2}$\;;
\item generate two numbers $\xi_{1}$, $\xi_{2}$ independently, each from the
      density $g(\xi)/\xi$, and $\vhi$ uniformly in $[0,2\pi)$\;;
\item put $z\lar(\xi_2-\xi_1)/(\xi_2+\xi_1)$,\; $q^0\lar E(\xi_2+\xi_1)$ and 
      $\qvec\lar q^0\Ro_p^{-1}\hat{n}(z,\vhi)$\;;
\item put $k\lar\Bo_{p_1+p_2}^{-1}q$\;;
\end{enumerate}
\end{Alg}

\section{A complete QCD antenna}
The straightforward way to generate $n$ momenta with the antenna structured 
density is by repeated use of the basic antenna. Let us denote 
\begin{equation}
      dA^i_{j,k}\df dA(q_j,q_k;q_i) \;\;,
\end{equation}
then 
\begin{equation}
   dA^2_{1,{\np}}dA^3_{2,{\np}}dA^4_{3,{\np}}\cdots 
            dA^{{\np}-1}_{{\np}-2,{\np}} 
   \;=\; \frac{\ipb{q_1}{q_{\np}}\,g_{\np}(\{q\}_{\np})}
              {\pi^{{\np}-2}\ipb{q_1}{q_2}\ipb{q_2}{q_3}\ipb{q_3}{q_4}\cdots
	                    \ipb{q_{\np-1}}{q_{\np}}}\,
          \prod_{i=2}^{{\np}-1}\dfq_i\vt(q_i) \;\;,\notag 
\end{equation}
where
\begin{equation}
   g_{\np}(\{q\}_{\np}) \;\df\; 
     g\left(\frac{\ipb{q_1}{q_2}}{\ipb{q_1}{q_{\np}}}\right)
     g\left(\frac{\ipb{q_2}{q_{\np}}}{\ipb{q_1}{q_{\np}}}\right)
     g\left(\frac{\ipb{q_2}{q_3}}{\ipb{q_2}{q_{\np}}}\right)
     g\left(\frac{\ipb{q_3}{q_{\np}}}{\ipb{q_2}{q_{\np}}}\right)\cdots
%     g\left(\frac{\ipb{q_{\np-2}}{q_{\np-1}}}{\ipb{q_{\np-2}}{q_{\np}}}\right)
     g\left(\frac{\ipb{q_{\np-1}}{q_{\np}}}{\ipb{q_{\np-2}}{q_{\np}}}\right)
	   \;\;.
\end{equation}
So if we have two momenta $q_1$ and $q_{\np}$, then we can easily generate
$\np-2$ momenta $q_j$ with the antenna structure. Remember that this
differential PS volume is completely invariant under Lorentz
transformations and scaling transformations, so that it seems self-evident to 
force the set of generated momenta in the CMF with a given energy, using the 
same kind of transformation as in the case of \rambo. 
If the first two momenta are
generated with density $f(q_1,q_{\np})$, then the UAF tells us that generated 
density $\Ant(\{p\}_{\np})$ satisfies
\begin{align}
   \Ant(\{p\}_{\np}) \;=\; 
   &\int \dfq_1\vt(q_1)\dfq_{\np}\vt(q_{\np})\,f(q_1,q_{\np})\,
    dA^2_{1,{\np}}dA^3_{2,{\np}}dA^4_{3,{\np}}\cdots 
            dA^{{\np}-1}_{{\np}-2,{\np}}\notag\\
       &\times d^4b\,\de^4(b-q_{(\np)}/\invm{q_{(\np)}})\,
        dx\,\de(x-\wcm/\invm{q_{(\np)}})\,\prod_{i=1}^{\np}\de^4(p_i-x\Bo_bq_i) 
	\;.
\end{align}
If we apply the same manipulations as in the proof of the correctness of
\rambo, we obtain the equation 
\begin{align}
   \Ant(\{p\}_{\np}) \;&=\; 
   \Theta_{w}(\{p\}_\np)\,
   \frac{\ipb{p_1}{p_{\np}}\,g_{\np}(\{p\}_{\np})}
        {\pi^{\np-2}\ipb{p_1}{p_2}\ipb{p_2}{p_3}\ipb{p_3}{p_4}\cdots
	 \ipb{p_{\np-1}}{p_{\np}}}
     \notag\\
       &\times\;\int d^4b\,\de(\invs{b}{}-1)\,dx\,\frac{2\scm^2}{x^5}\,
                f(x^{-1}\Bo_b^{-1}p_1\,,\,x^{-1}\Bo_b^{-1}p_{\np})  \;\;.
\label{SarEq003}	       
\end{align}
Now we choose $f$ such that $q_1$ and $q_{\np}$ 
are generated back-to-back in their CMF with total energy $\wcm$, i.e.,
\begin{equation}
   f(q_1,q_{\np}) \;=\; \frac{2}{\pi}\,\de^4(q_1+q_{\np}-\wcm\enul) \;\;.
\end{equation}
If we evaluate the second line of \eqn{SarEq003} with this $f$, we arrive at 
\begin{multline}
   \frac{4\scm^2}{\pi}\int dx\,\frac{1}{x^5}\,d^4b\,\de(\invs{b}{}-1)\,
   \de^4(x^{-1}\Bo_b^{-1}(p_1+p_{\np})-\wcm\enul)  \\
   \;=\; \frac{4}{\pi}\int_0^\infty dx\,\frac{1}{x^5}\,
          \de\left(\frac{\invs{(p_1+p_{\np})}{}}{\scm x^2}-1\right)
    \;=\; \frac{\scm^2}{2\pi\ipb{p_1}{p_{\np}}^{2}} \;\;,	  
\end{multline}
so that the generated density is given by  
\begin{align}
  \Ant(\{p\}_n) \;=\; 
    \Theta_{\scm}(\{p\}_\np)\,
    \frac{\scm^2}
         {2\pi^{\np-2}}\,
    \frac{g_{\np}(\{p\}_{\np})}
         {\ipb{p_1}{p_2}\ipb{p_2}{p_3}\ipb{p_3}{p_4}\cdots
	  \ipb{p_{\np-1}}{p_{\np}}\ipb{p_{\np}}{p_1}}   \;\;.  
\label{SarEq010}		      
\end{align}
Note that, somewhat surprisingly, also the factor $\ipb{p_{\np}}{p_1}^{-1}$
comes out, thereby making the antenna even more symmetric. In fact, if the 
density $f(q_1,q_2)=c^4\exp(-cq_1^0-cq_2^0)/4\pi^2$ is taken instead of the one
we just used, the calculation can again be done exactly, with exactly the same 
result. The algorithm to generate $n$ momenta with the above antenna structure
is given by 
\begin{Alg}[\texttt{QCD ANTENNA}]
\begin{enumerate}
\item generate massless momenta $q_1$ and $q_{\np}$;
\item generate $n-2$ momenta $q_j$  by the basic
      antennas $dA^2_{1,{\np}}dA^3_{2,{\np}}dA^4_{3,{\np}}\cdots 
                dA^{{\np}-1}_{{\np}-2,{\np}}$;
\item compute $q_{(\np)} = \sum_{j=1}^{\np}q_j$, and the
      boost and scaling transforms that bring $q_{(\np)}$ to $\wcm\enul$;
\item for $j=1,\ldots,{\np}$, boost and scale the $q_j$ accordingly, into the 
      $p_j$.
\end{enumerate}
\end{Alg}
Usually, the event generator is used to generate cut PS. 
If a generated event does not satisfy the physical cuts, it is rejected. In the
calculation of the weight coming with an event, the only contribution coming
from the functions $g$ is, therefore, their normalization. In total, this gives
a factor $1/(2\log\xim)^{2{\np}-4}$ in the density.

\section{Incoming momenta and symmetrization}
The density given by the algorithm above, is not quite what we want. First of
all, we want to include the incoming momenta $\ppin$ and $\pnip$ in the APS, so
that the density becomes proportional to $[\ipb{\ppin}{p_1}\ipb{p_1}{p_2}\cdots
\ipb{p_{\np-1}}{p_{\np}}\ipb{p_{\np}}{\pnip}]^{-1}$ instead of
$[\ipb{p_1}{p_2}\cdots\ipb{p_{\np-1}}{p_{\np}}\ipb{p_{\np}}{p_1}]^{-1}$. Then
we want the sum of all permutations of the momenta, including the incoming
ones.

\subsection{Generating incoming momenta\label{SecInc}}
The incoming momenta can be generated after the antenna has been generated. 
To show how, let us introduce the following ``regularized'' scalar product:
\begin{equation}
   \ipb{p}{q}_\de \;\df\;    \ipb{p}{q} + \de p^0q^0\;\;,
\end{equation}
where $\de$ is a small positive number. This regularization is not completely
Lorentz invariant, but that does not matter here. Important is that it is still
invariant under rotations, as we shall see. Using this regularization, we are
able to generate a momentum $k$ with a probability density
\begin{equation}
   \frac{1}{2\pi I_\de(p_1,p_2)}\,
   \frac{\vt(k)\,\de(k^0-1)}{\ipb{p_1}{k}_\de\ipb{\tilde{k}}{p_2}_\de} \;\;.
\label{SarEq004}       
\end{equation}
To show how, we calculate the normalization $I_\de(p_1,p_2)$. Using the 
Feynman-representation of $1/[\ipb{p_1}{k}_\de\ipb{\tilde{k}}{p_2}_\de]$, it 
is easy to see that
\begin{equation}
   I_\de(p_1,p_2)
   \;=\; \frac{1}{4\pi p_1^0p_2^0}\int dzd\vhi\int_0^1
         \frac{dx}{(1+\de-|\pvec_x|z)^2} \;\;,
\end{equation}
where $\pvec_x=x\hat{p}_1+(x-1)\hat{p}_2$. The integral over $z$ and $\vhi$
can now be performed, with the result that
\begin{equation}
   I_\de(p_1,p_2)
   \;=\; \frac{1}{p_1^0p_2^0}\int_0^1\frac{dx}{(1+\de)^2-|\pvec_x|^2}
   \;=\; \frac{1}{2\ipb{p_1}{\tilde{p}_2}}
         \int_0^1\frac{dx}{(x_+-x)(x-x_-)} \;\;,
\end{equation}
where $x_{\pm}$ are the solutions for $x$ of the equation 
$1+\de=|\pvec_x|$. Further evaluation finally leads to 
\begin{equation}
   I_\de(p_1,p_2) 
   \;=\; \frac{\ipb{p_1}{\tilde{p}_2}^{-1}}{x_+-x_-}\,
         \log\left|\frac{x_+}{x_-}\right| \;\;,\quad
   x_\pm 
   \;=\; \frac{1}{2}\pm\frac{1}{2}
         \sqrt{1+\frac{2p_1^0p_2^0(2\de+\de^2)}{\ipb{p_1}{\tilde{p_2}}}} \;\;.
\end{equation}
Notice that there is a smooth limit to the case in which $p_1$ and $p_2$ are 
back-to-back:
\begin{equation}
   I_\de(p,\tilde{p}) \;=\; \lim_{q\ra\tilde{p}}I_\de(p,q) 
   \;=\;\frac{1}{(p^0)^2(2\de+\de^2)} \;\;.
\end{equation}
The algorithm to generate $k$ can be derived by reading the 
evaluations of the integrals backwards.

Because $k$ and
$\tilde{k}$ are back-to-back, they can serve as the incoming momenta. To fix
them to $\enul+ \ethr$ and $\enul-\ethr$, the whole system of momenta can be
rotated. If we generate momenta with the density $\Ant$, use the first 
two momenta to generate the incoming momenta and rotate, we get a density
\begin{align}
  D_{\scm}(\{p\}_{\np})
   \;&=\;\int d^{4n}q\,\Ant(\{q\}_n)\;\dfk\,
         \frac{1}{2\pi I_\de(q_1,q_2)}
         \frac{\vt(k)\,\de(k^0-1)}
              {\ipb{q_1}{k}_\de\ipb{q_2}{\tilde{k}}_\de}\,   
         \prod_{i=1}^{\np}\de^4(p_i-\Ro_{k}q_i)      \notag\\
     &=\;  \Ant(\{p\}_n)\,I_\de(p_1,p_2)^{-1}\, 
           \int \dfk\vt(k)\,\de(k^0-1)\,
	   \frac{(2\pi)^{-1}}
	        {\ipb{p_1}{\Ro_{k}k}_\de\ipb{p_2}{\Ro_{k}\tilde{k}}_\de} \;\;, 
\end{align}
where we used the fact that the whole expression is invariant under rotations, 
and that these are orthogonal transformations.
The last line of the previous expression can be evaluated further with the 
result that
\begin{equation}
   D_{\scm}(\{p\}_{\np})
   \;=\; \Ant(\{p\}_{\np})\,\frac{I_\de(p_1,p_2)^{-1}}
	        {\ipb{p_1}{\ppin}_\de\ipb{\pnip}{p_2}_\de}
   \quad\textrm{with}\quad
   \ppin=\enul+\ethr \;,\;\; \pnip=\enul-\ethr \;.
\label{SarEq012}		
\end{equation}
The algorithm to generate the incoming momenta is given by
\begin{Alg}[\texttt{INCOMING MOMENTA}]\label{SarAlg1}
\begin{enumerate}
\item given a pair $\{p_1,p_2\}$, calculate $x_+$ and $x_-$; 
\item generate $x$ in $[0,1]$ with density $\sim [(x_+-x)(x-x_-)]^{-1}$, 
      and put $\pvec_x\lar x\hat{p}_1+(x-1)\hat{p}_2$\;;
\item generate $\vhi$ uniformly in $[0,2\pi)$, $z$ in $[-1,1]$ with 
      density $\sim(1+\de-|\pvec_x|z)^{-2}$\;;
\item put $\vec{k}\lar\Ro_{p_x}^{-1}\hat{n}(z,\vhi)$
      and $k^0\lar1$\;;       
\item rotate all momenta with $\Ro_{k}$\;;      
\item put $\ppin\lar\half\wcm(e_0+e_3)$ and 
          $\pnip\lar\half\wcm(e_0-e_3)$\;.      
\end{enumerate}
\end{Alg}
Notice that $I_\de(p_1,p_2)\ipb{p_1}{\ppin}_\de\ipb{\pnip}{p_2}_\de$ is
invariant under the scaling $p_1,p_2\ra cp_1, cp_2$ with a constant $c$, so
that scaling of $\ppin$ and $\pnip$ has no influence on the density.

The pair $(q_1,q_2)$ with which $k$ is generated is free to choose because we
want to symmetrize in the end anyway. We should only choose it such, that we
get rid of the factor $\ipb{q_1}{q_2}$ in the denominator of
$\Ant(\{q\}_{\np})$.

\subsection{Choosing the type of antenna with incoming momenta\label{SecCho}}
A density which is the sum over permutations can be obtained by generating
random permutations, and returning the generated momenta with permutated
labels. This, however, only makes sense for the outgoing momenta. The incoming
momenta are fixed, and should be returned separately from the outgoing momenta
by the event generator. Therefore, a part of the permutations has to be
generated explicitly. There are two kinds of terms in the sum: those in which
$\ipb{\ppin}{\pnip}$ appears, and those in which it does not.

\paragraph*{Case 1: antenna with $\ipb{\ppin}{\pnip}$.}
To generate the first kind, we can choose a label $i$ at random with weight
$\ipb{p_i}{p_{i+1}}/\Sigma_1(\{p\}_{\np})$
where $\Sigma_1(\{p\}_{\np})$ is the sum of all scalar products in the 
antenna~\footnote{Read $i+1\!\mod\np$ when $i+1$ occurs in this section}:
\begin{equation}
   \Sigma_1(\{p\}_{\np}) \;\df\; \sum_{i=1}^n\ipb{p_i}{p_{i+1}}  \;\;.
\end{equation}
This is a proper weight, since all scalar products are positive. The total
density gets this extra factor then, so that $\ipb{p_i}{p_{i+1}}$ cancels. The
denominator of the weight factor does not give a problem, because its singular
structure is much softer than the one of the antenna. The pair
$\{p_i,p_{i+1}\}$ can then be used to generate the incoming momenta, as shown
above. So in this case, a density 
$\Ant(\{p\}_{\np})B_1(\{p\}_{\np})/\Sigma_1(\{p\}_{\np})$ is generated, where
\begin{equation}
   B_1(\{p\}_{\np}) \;\df\; 
   \sum_{i=1}^{\np}\frac{\ipb{p_i}{p_{i+1}}\,I_\de(p_i,p_{i+1})^{-1}}
        {\ipb{p_i}{\ppin}_\de\ipb{\pnip}{p_{i+1}}_\de} \;\;.
\end{equation}

\paragraph*{Case 2: antenna without $\ipb{\ppin}{\pnip}$.}
To generate the second kind, we can choose two non-equal labels $i$ and
$j$ with weight $\ipb{p_i}{p_{i+1}}\ipb{p_j}{p_{j+1}}/\Sigma_2(\{p\}_{\np})$, 
where 
\begin{equation}
   \Sigma_2(\{p\}_{\np})
   \;\df\;\sum_{i\neq j}^{\np}\ipb{p_i}{p_{i+1}}\ipb{p_j}{p_{j+1}} \;\;.
\end{equation}
Next, a pair $(k,l)$ 
of labels has to be chosen from the set  of pairs
\begin{equation}
   \{(i,j)\}_+
   \;\df\; \{(i,j)\,,\,(i,{j+1})\,,\,({i+1},j)\,,\,({i+1},{j+1})\} \;\;.
\label{SarEq019}   
\end{equation}
If this is done with weight $I_\de(p_k,p_l)/\Sigma_{i,j}(\{p\}_{\np})$, where
\begin{equation}
   \Sigma_{i,j}(\{p\}_{\np})
   \;\df\;  \sum_{(k,l)\in\{(i,j)\}_+}I_\de(p_k,p_l)  \;\;,
\end{equation}
then the density $\Ant(\{p\}_{\np})B_2(\{p\}_{\np})/\Sigma_2(\{p\}_{\np})$ is
generated, where
\begin{align}
  B_2(\{p\}_{\np}) 
  \;&=\; \sum_{i\neq j}^\np\ipb{p_i}{p_{i+1}}\ipb{p_j}{p_{j+1}}
        \sum_{(k,l)\in\{(i,j)\}_+}\frac{I_\de(p_k,p_l)}
                                  {\Sigma_{i,j}(\{p\}_{\np})}\cdot
                             \frac{I_\de(p_k,p_l)^{-1}}
			          {\ipb{p_k}{\ppin}_\de\ipb{\pnip}{p_l}_\de}
				  \notag\\
  \;&=\; \sum_{i\neq j}^{\np}
      \frac{\ipb{p_i}{p_{i+1}}\ipb{p_j}{p_{j+1}}}
           {\ipb{p_i}{\ppin}_\de\ipb{p_{i+1}}{\ppin}_\de
	    \ipb{\pnip}{p_j}_\de\ipb{\pnip}{p_{j+1}}_\de}\cdot
	    \frac{\sum_{(k,l)\in\{(i,j)\}_+}\ipb{p_k}{\ppin}_\de
	                                    \ipb{\pnip}{p_l}_\de}
	         {\sum_{(k,l)\in\{(i,j)\}_+}I_\de(p_k,p_l)} \;\;.
\end{align}
Before all this, we first have to choose between the two cases, and the 
natural way to do this is with relative weights 
$\sfrac{1}{2}\scm\Sigma_1(\{p\}_{\np})$ and $\Sigma_2(\{p\}_{\np})$, 
so that the complete density is equal to 
\begin{align}
   S^{\scriptscriptstyle\textrm{QCD}}_{\scm}(\{p\}_{\np}) \;=\;
   \frac{1}{n!}\sum_{\textrm{perm.}}
   \Ant(\{p\}_{\np})\,
   \frac{\sfrac{1}{2}\scm B_1(\{p\}_{\np})+B_2(\{p\}_{\np})}
        {\sfrac{1}{2}\scm\Sigma_1(\{p\}_{\np})+\Sigma_2(\{p\}_{\np})} \;\;,
\label{SarEq009}		     
\end{align}
where the first sum is over all permutations of $(1,\ldots,n)$. One can, of
course, try to optimize the weights for the two cases using the adaptive
multichannel method (cf. \cite{Pittau1}). The result of using the sum of the
two densities is that the factors $\ipb{p_i}{p_{i+1}}$ in the numerator of
$B_1(\{p\}_{\np})$ and $\ipb{p_i}{p_{i+1}}\ipb{p_j}{p_{j+1}}$ in the numerator
of $B_2(\{p\}_{\np})$ cancel with the same factors in the denominator of
$\Ant(\{p\}_\np)$, so that we get exactly the pole structure we want. The
`unwanted' singularities in $B_1(\{p\}_{\np}),B_2(\{p\}_{\np})$ and
$\Sigma_1(\{p\}_{\np}),\Sigma_2(\{p\}_{\np})$ are much softer than the ones
remaining in $\Ant(\{p\}_\np)$, and cause to trouble.  The algorithm to
generate the incoming momenta and the permutation is given by
\begin{Alg}[\texttt{CHOOSE INCOMING POLE STRUCTURE}]
\begin{enumerate}
\item choose case 1 or 2 with relative weights 
      $\sfrac{1}{2}\scm\Sigma_1(\{p\}_{\np})$ and 
      $\Sigma_2(\{p\}_{\np})$\;;
\item in case 1, choose $i_1$ with relative weight 
      $\ipb{p_{i_1}}{p_{i_1+1}}$ and put $i_2\lar i_1+1$\;;
\item in case 2, choose $(i,j)$ with $(i\neq j)$ and relative weight 
      $\ipb{p_i}{p_{i+1}}\ipb{p_j}{p_{j+1}}$, and then \\
      choose $(i_1,i_2)$ 
      from $\{(i,j)\}_+$ with 
      relative weight $I_\de(p_{i_1},p_{i_2})$\;;
\item use $\{p_{i_1},p_{i_2}\}$ to generate the incoming momenta with 
      Algorithm \ref{SarAlg1};
\item generate a random permutation $\si\in S_{n}$ and put
      $p_i\leftarrow p_{\si(i)}$ for all $i=1,\ldots,n$. 
\end{enumerate}
\end{Alg}
An algorithm to generate the random permutations can be found in \cite{Knuth}.
An efficient 
algorithm to calculate a sum over permutations can be found in \cite{Kuijf1}.

\section{Improvements}
When doing calculations with this algorithm on a PS, cut such that
$\invs{(p_i+p_j)}{}>s_0$ for all $i\neq j$ and some reasonable $s_0>0$, we
notice that a very high percentage of the generated events does not pass the
cuts.  An important reason why this happens is that the cuts, generated by the
choices of $g$ (\eqn{SarEq006}) and $\xim$ (\eqn{SarEq007}), are implemented
only on quotients of scalar products that appear explicitly in the generation
of the QCD-antenna:
\begin{equation}
   \xi^i_1 \df \frac{\ipb{p_{i-1}}{p_i}}{\ipb{p_{i-1}}{p_\np}} 
   \qquad\textrm{and}\qquad
   \xi^i_2 \df \frac{\ipb{p_{i}}{p_{\np}}}{\ipb{p_{i-1}}{p_\np}}  \;\;,
   \quad i=2,3\ldots,\np-1\;\;.
\end{equation}
The total number of these $\xi$-variables is
\begin{equation}
   n_\xi \df 2\np-4 \;\;,
\end{equation}
and the cuts are implemented such that 
$\xim^{-1}\leq\xi^i_{1,2}\leq\xim$ for $i=2,3\ldots,\np-1$.
We show now how these cuts can be implemented on {\em all} quotients 
\begin{equation}
   \frac{\ipb{p_{i-1}}{p_{i}}}{\ipb{p_{j-1}}{p_{j}}} \;,\quad
   \frac{\ipb{p_{i-1}}{p_{i}}}{\ipb{p_{j}}{p_{\np}}} \quad\textrm{and}\quad
   \frac{\ipb{p_{i}}{p_{\np}}}{\ipb{p_{j}}{p_{\np}}} \;,
   \quad i,j=2,3,\ldots,\np-1 \;\;.
\label{SarEq013}   
\end{equation}
We define the $m$-dimensional convex polytope 
\begin{equation}
   \Pol_m\df\{(x_1,\ldots,x_m)\in[-1,1]^m\,\big|\;
                  |x_i-x_j|\leq1\;\forall\,i,j=1,\ldots,m\} \;\;,
\label{SarEq015}		  
\end{equation}
and replace the generation of the the $\xi$-variables by the following:
\begin{Alg}[\texttt{IMPROVEMENT}]
\begin{enumerate}
\item generate $(x_1,x_2,\ldots,x_{n_\xi})$ distributed uniformly in 
      $\Pol_{n_\xi}$;
\item define $x_0\df0$ and put,  
      \begin{equation}
         \xi^i_1\lar e^{(x_{2i-3}-x_{2i-4})\log\xim}\;\;,\quad
	 \xi^i_2\lar e^{(x_{2i-2}-x_{2i-4})\log\xim}
	 \label{SarEq011}
      \end{equation}
      for all $i=2,\ldots,n-1$.
\end{enumerate}
\end{Alg}
Because all the variables $x_i$ are distributed uniformly such that
$|x_i-x_j|\leq1$, {\em all} quotients of (\ref{SarEq013}) are distributed such
that they are between $\xim^{-1}$ and $\xim$. In terms of the variables $x_i$,
this means that we generate the volume of $\Pol_{n_\xi}$, which is $n_\xi+1$,
instead of the volume of $[-1,1]^{n_\xi}$, which is $2^{n_\xi}$.  
In \cite{HKPol}, 
we give the algorithm to generate variables distributed uniformly in
$\Pol_m$. We have to note here that this improvement only makes sense because
the algorithm to generate these variables is very efficient. The total density
changes such, that the function $g_n$ in \eqn{SarEq010} has to be replaced by
\begin{equation}
   g^{\Pol}_{n}(\xim;\{\xi\})
   \;\df\; \frac{1}{(n_\xi+1)(\log\xim)^{n_\xi}}\,
           \theta(\,(x_1,\ldots,x_{n_\xi})\in\Pol_{n_\xi}\,) \;\;,
\end{equation}
where the variables $x_i$ are functions of the variables $\xi^i_{1,2}$ as
defined by $(\ref{SarEq011})$.  Because hit-and-miss MC is used to restrict
generated events to cut PS, again only the normalization has to be calculated
for the weight of an event.

With this improvement, still a large number of events does not pass the cuts.
% When we made plots of surfaces in $\Pol_{n_\xi}$, we noticed that the corners
% of $\Pol_{n_\xi}$ are less populated by accepted events, than would be 
% expected from a uniform distribution. This inspired us to put a non-uniform
% distribution around the center of $\Pol_{n_\xi}$. If it is chosen to be
% proportional to $\prod_{i=1}^{n_\xi}\cos(\frac{\pi}{2}x_i)$, the generated
% volume can still be calculated exactly and there is an efficient algorithm. 
% We tried this, and the percentage of accepted events was higher,but following
% method appeared to be more succesfull.
% 
The situation with PS is depicted in \fig{SarFig02}.
\begin{figure}
\begin{center}
\begin{picture}(160,100)(0,0)
\GOval(80,50)(80,50)(90){1}
\GOval(80,50)(56,35)(90){0.7}
\BBoxc(80,50)(72,50)
\Line(133,75)(170,85)\Text(175,86)[l]{phase space}
\Line(90,40)(170,20)\Text(175,20)[l]{cut phase space}
\Line(35,50)(-20,50)\Text(-25,51)[r]{generated phase space}
\end{picture}
\Cap{Schematic view on phase space.}
\label{SarFig02}
\end{center}
\vspace{-20pt}
\end{figure}
Phase space contains generated phase space which contains cut phase space.  The
problem is that most events fall in the shaded area, which is the piece of
generated PS that is not contained in cut PS. To get a higher
percentage of accepted events, we use a random variable $\xi_v\in[0,\xim]$,
instead of the fixed number $\xim$, to generate the variables
$\xi^{i}_{1,2}$. This means that the size of the generated PS
becomes variable.  If this is done with a probability distribution such that
$\xi_v$ can, in principle, become equal to $\xim$, then whole of cut phase
space is still covered. We suggest the following, tunable, density:
\begin{equation}
   h_\al(\xi_v) 
   \;=\; \frac{\al n_\xi+1}{(\log\xim)^{\al n_\xi+1}}\cdot
         \frac{(\log\xi_v)^{\al n_\xi}}{\xi_v}\,\theta(1\leq\xi_v\leq\xim) 
	 \;\;,\quad \al\geq0 \;\;.
\end{equation}
If $\al=0$, then $\log\xi_v$ is distributed uniformly in $[0,\log\xim]$, and
for larger $\al$, the distribution peaks more and more towards $\xi_v=\xim$.
Furthermore, the variable is easy to generate and the total generated density
can be calculated exactly: 
$g^{\Pol}_{n}(\xim;\{\xi\})$ should be replaced by 
\begin{align}
   G^{\Pol}_{n}(\al,\xim;\{\xi\}) 
   \;\df&\;\int d\xi_v\,h_\al(\xi_v)\,g^{\Pol}_{n}(\xi_v;\{\xi\}) \notag\\
   \;=&\; \frac{1}{n_\xi+1}\cdot
          \frac{\al n_\xi+1}{(\log\xim)^{\al n_\xi+1}}
	  \int_{\log\xilow}^{\log\xim}dx\,x^{(\al-1)n_\xi} \;\;,
%         \cdot\frac{(\log\xim)^{\bar{\al}n_\xi+1}
%	            -(\log\xilow)^{\bar{\al}n_\xi+1}}{\bar{\al}n_\xi+1} \;\;,
\label{SarEq016}
\end{align}
where $\xilow$ is the maximum of the ratios of scalar products in 
(\ref{SarEq013}).
%\begin{equation}
%   \xilow\df\max_{\{i,j\}}\xi^{i}_{j} \;\;.
%\end{equation}
%If $\bar{\al}n_\xi+1=0$, then the last factor should be replaced by 
%$\log\log\xim - \log\log\xilow$.
% The total density with which the momenta are generated is now given by 
% \eqn{SarEq009}, with $\Ant(\{p\}_{\np})$ from (\eqn{SarEq010}) replaced by 
% \begin{equation}
%    \Ant_{n,\Pol}(\al;\{p\})
%    \;\df\; \frac{1}{\ipb{1}{2}\ipb{2}{3}\ipb{3}{4}\cdots
%            \ipb{n-1}{n}\ipb{n}{1}}\cdot 
%            \frac{\scm^2}{2\pi^{n-1}}\,G^{\Pol}_{n}(\al,\xim;\{\xi\}) \;\;. 
% \end{equation}
% We cannot put the cosine-distribution in $\Pol_{n_\xi}$ now, because then 
% $G^{\Pol}_{n}(\al,\xim;\{\xi\})$ cannot be calculated exactly. 

\section{Results and conclusions}
We compare \sarge\ with \rambo\ in the integration of the {\tt SPHEL}-integrand
for processes of the kind $\gl\gl\ra \np\gl$, which is given by 
\begin{equation}
   \sum_{\textrm{perm.}}
   \frac{2\sum_{i\neq j}^{n+1}
   \ipb{p_i}{p_j}^4}{\ipb{p_1}{p_2}\ipb{p_2}{p_3}\ipb{p_3}{p_4}\cdots
            \ipb{p_{\np}}{p_{\np+1}}\ipb{p_{\np+1}}{p_{\np+2}}
	                            \ipb{p_{\np+2}}{p_1}} \;\;,
\end{equation}
where $p_1$ and $p_2$ are the incoming momenta, and the first sum is over all 
permutations of $(2,3,\ldots,n+2)$ except the cyclic permutations.
The results are presented in \tab{SarTab02}.
\begin{table}[b]
\begin{center}
\begin{tabular}{|>{$}c<{$}|}
  \hline n \\
  \hline \tau_{\texttt{SPHEL}}(\seco) \\ 
  \hline \tau_{\textrm{exact}}(\seco) \\ \hline
\end{tabular}  
\begin{tabular}{|>{$}c<{$}|>{$}c<{$}|>{$}c<{$}|>{$}c<{$}|}
  \hline 4 &5 &6 &7 \\
  \hline 5.40\emu{5} &2.70\emu{4} &1.80\emu{3} &1.41\emu{2} \\ 
  \hline 3.07\emu{1} &1.08 &3.35 &10.92 \\ \hline
\end{tabular}  
\end{center}
\vspace{-10pt}
\Cap{cpu-times ($\tau_{\texttt{SPHEL}}$) in seconds needed to evaluate the 
         {\tt SPHEL}-integrand one time with a $300$-MHz UltraSPARC-IIi 
	 processor, and the cpu-times ($\tau_{\textrm{exact}}$) needed to 
	 evaluate the exact integrand, estimated with the help of 
	 \tab{SarTab01}.}  
\label{SarTab03}	 
\end{table}
\begin{table}
\begin{center}
% \begin{tabular}{|>{$}c<{$}|}
%    \hline \wcm=1000,\quad p_{T}=40,\quad\theta_0=30^\circ \\ 
%    \hline
% \end{tabular}\\
\begin{tabular}{|c|}
   \hline \\ $\gl\gl\rightarrow4\gl$ \\ \\ $1\%$ error \\ \\ \hline
\end{tabular}
\begin{tabular}{|c|}
   \hline alg.     \\ 
   \hline $\sigma$ \\ 
   \hline $\Nge$   \\
   \hline $\Nac$   \\
   \hline $\tcpu(\hour)$  \\ 
   \hline $\texa(\hour)$  \\ \hline
\end{tabular}
\begin{tabular}{|>{$}c<{$}|>{$}c<{$}|>{$}c<{$}|>{$}c<{$}|}
  \hline \rambo     &\sarge,\al=0.0 &\sarge,\al=0.5 &\sarge,\al=10.0\\
  \hline 4.30\ee{8} &4.31\ee{8}     &4.37\ee{8}     &4.32\ee{8} \\
  \hline 4,736,672  &296,050        &278,702        &750,816 \\
  \hline 3,065,227  &111,320        &40,910         &23,373 \\
  \hline 0.198      &0.0254         &0.0172         &0.0348 \\
  \hline 262        &9.52           &3.51           &2.03 \\ \hline
\end{tabular}

%\end{center}
%\begin{center}
\begin{tabular}{|c|}
   \hline \\ $\gl\gl\rightarrow5\gl$ \\ \\ $1\%$ error \\ \\ \hline
\end{tabular}
\begin{tabular}{|c|}
   \hline alg.     \\ 
   \hline $\sigma$ \\ 
   \hline $\Nge$   \\
   \hline $\Nac$   \\
   \hline $\tcpu(\hour)$  \\ 
   \hline $\texa(\hour)$  \\ \hline
\end{tabular}
\begin{tabular}{|>{$}c<{$}|>{$}c<{$}|>{$}c<{$}|>{$}c<{$}|}
  \hline\rambo       &\sarge,\al=0.0 &\sarge,\al=0.5 &\sarge,\al=10.0\\
  \hline 3.78\ee{10} &3.81\ee{10}    &3.80\ee{10}    &3.81\ee{10} \\
  \hline 4,243,360   &715,585        &1,078,129      &6,119,125 \\
  \hline 1,712,518   &167,540        &36,385         &21,111 \\
  \hline 0.286       &0.133          &0.0758         &0.277 \\
  \hline 514         &51.6           &11.7           &9.10 \\ \hline
\end{tabular}

%\end{center}
%\begin{center}
\begin{tabular}{|c|}
   \hline \\ $\gl\gl\rightarrow6\gl$ \\ \\ $1\%$ error \\ \\ \hline
\end{tabular}
\begin{tabular}{|c|}
   \hline alg.     \\ 
   \hline $\sigma$ \\ 
   \hline $\Nge$   \\
   \hline $\Nac$   \\
   \hline $\tcpu(\hour)$  \\ 
   \hline $\texa(\hour)$  \\ \hline
\end{tabular}
\begin{tabular}{|>{$}c<{$}|>{$}c<{$}|>{$}c<{$}|>{$}c<{$}|}
  \hline \rambo      &\sarge,\al=0.0 &\sarge,\al=0.5 &\sarge,\al=10.0\\
  \hline 3.07\ee{12} &3.05\ee{12}    &3.13\ee{12}    &3.05\ee{12} \\
  \hline 3,423,981   &2,107,743      &6,136,375      &68,547,518 \\
  \hline 700,482     &276,344        &34,095         &17,973 \\
  \hline 0.685       &1.32           &0.471          &3.17 \\
  \hline 653         &258            &32.2           &19.9 \\ \hline
\end{tabular}

%\end{center}
%\begin{center}
\begin{tabular}{|c|}
   \hline \\ $\gl\gl\rightarrow7\gl$ \\ \\ $3\%$ error \\ \\ \hline
\end{tabular}
\begin{tabular}{|c|}
   \hline alg.     \\ 
   \hline $\sigma$ \\ 
   \hline $\Nge$   \\
   \hline $\Nac$   \\
   \hline $\tcpu(\hour)$  \\ 
   \hline $\texa(\hour)$  \\ \hline
\end{tabular}
\begin{tabular}{|>{$}c<{$}|>{$}c<{$}|>{$}c<{$}|>{$}c<{$}|}
  \hline \rambo      &\sarge,\al=0.0 &\sarge,\al=0.5 &\sarge,\al=10.0\\    
  \hline 2.32\ee{14} &2.16\ee{14}    &2.20\ee{14}    &2.28\ee{14} \\
  \hline 605,514     &710,602        &5,078,153      &125,471,887 \\
  \hline 49,915      &42,394         &3,256          &1,789 \\
  \hline 0.224       &1.86           &0.452          &6.74 \\
  \hline 152         &130            &10.3           &12.2 \\ \hline
\end{tabular}
\caption{Results for the integration of the {\tt SPHEL}-integrand.
         The CM-energy and the cuts used are 
	 $\wcm=1000$, $p_{T}=40$ and $\theta_0=30^\circ$. Presented are 
	 the finial result ($\si$), the number of generated ($\Nge$) and 
	 accepted ($\Nac$) events, the cpu-time $(\tcpu)$ in hours, and
	 the cpu-time ($\texa$) it would take to integrate the exact matrix 
	 element, estimated with the help of \tab{SarTab03}.
	 In the calculation of this table, adaptive multichanneling 
	 in the two cases of \Sec{SecCho} was used, and $\delta=0.01$ 
	 (\Sec{SecInc}).}
\label{SarTab02}	 
\end{center}
\end{table}
\begin{figure}
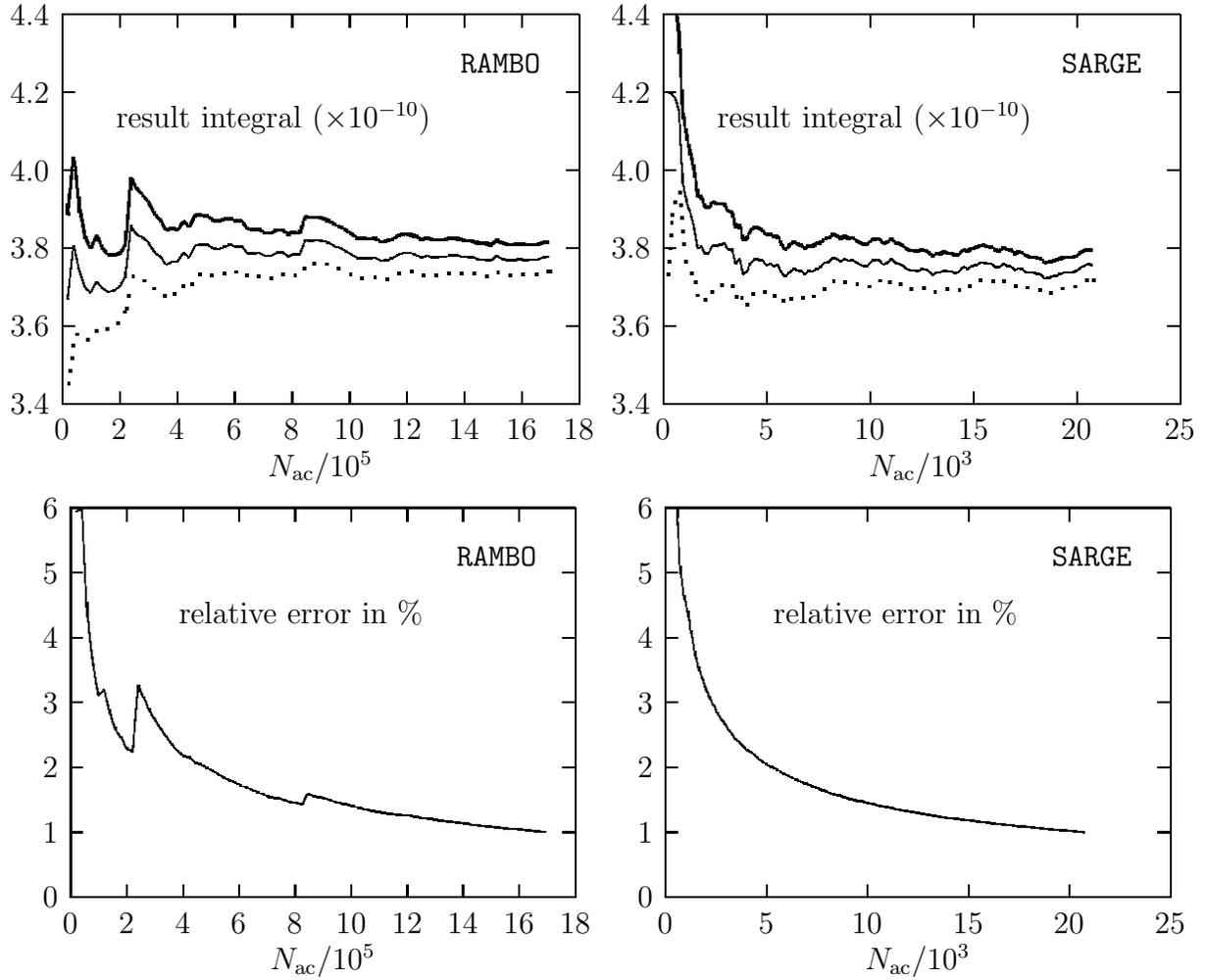

\begin{center}
\hspace{-45pt}
%
%
% GNUPLOT: LaTeX picture
\setlength{\unitlength}{0.240900pt}
\ifx\plotpoint\undefined\newsavebox{\plotpoint}\fi
\sbox{\plotpoint}{\rule[-0.200pt]{0.400pt}{0.400pt}}%

\caption{The convergence of the MC-process in the integration of the 
         {\tt SPHEL}-integrand for $\np=5$, with $\wcm=1000$, $p_T=40$ and
	 $\theta_0=30^\circ$. The upper graphs show the 
	 integral itself as function of the number of 
	 accepted events, together with the estimated bounds on the expected
	 deviations. The lower graphs show the relative
	 error. \sarge\ was used with adaptive multi-channeling in the two 
	 cases of \Sec{SecCho}, with $\delta=0.01$ (\Sec{SecInc}) and without 
	 the variable $\xi_v$. The number of generated events was $6,699,944$, 
	 and the cpu-time was $0.308$ hours.}
\label{SarFig03}	 
\end{center}
\end{figure}
The calculations were done at a CM-energy $\wcm=1000$ with cuts $p_T=40$ on
each transverse momentum and $\theta_0=30^\circ$ on the angles between the
momenta. We present the results for $\np=4,5,6,7$, calculated with \rambo\ and
\sarge\ with different values for $\al$ (\eqn{SarEq016}). The value of $\si$ is
the estimate of the integral at an estimated error of $1\%$ for $\np=4,5,6$ and
$3\%$ for $\np=7$. These numbers are only printed to show that different
results are compatible. Remember that they are not the whole cross sections: 
flux factors, color factors, sums and averages over helicities, and coupling 
constants are not included.
The other data are the number of generated events
($\Nge$), the number of accepted events ($\Nac$) that passed the cuts, the
cpu-time consumed ($\tcpu$), and the cpu-time the calculation would have
consumed if the exact matrix element had been used ($\texa$), both in hours.
This final value is estimated with the help of \tab{SarTab03} and the formula
\begin{equation}
   \texa 
   \;=\; \tcpu + \Nac(\tau_{\textrm{exact}} - \tau_{\texttt{SPHEL}})\;\;,
\end{equation}
where $\tau_{\textrm{exact}}$ and $\tau_{\texttt{SPHEL}}$ are the cpu-times it 
takes to evaluate the squared matrix element once.
Remember that the integrand only has to be evaluated for accepted events. The
calculations have been performed with a single $300$-MHz UltraSPARC-IIi 
processor.

The first conclusion we can draw is that \sarge\ outperforms \rambo\ in
computing time for all processes. This is especially striking for lower number
of outgoing momenta, and this behavior has a simple explanation: we kept the
CM-energy and the cuts fixed, so that there is less energy to distribute over
the momenta if $\np$ is larger, and the cuts become relatively tighter. As a
result, \rambo\ gains on \sarge\ if $\np$ becomes larger. This effect would 
not appear if the energy, or the cuts, would scale with $\np$ like in 
\tab{SarTab01}. 
Another indication for this effect is the fact that the ratio 
$\Nac/\Nge$ for \rambo, 
which estimates the ratio of the 
volumes of cut PS and whole PS,  
decreases with $\np$.

Another conclusion that can be drawn is that \sarge\ performs better if $\al$
is larger. Notice that the limit of $\al\ra\infty$ is equivalent with dropping
the improvement of the algorithm using the variable $\xi_v$ (\eqn{SarEq016}).
Only if the integrand becomes too flat, as in the case of $\np=7$ with the 
energy and the cuts as given in the table, smaller values are preferable. Then, 
too many events do not pass the cuts if $\alpha$ is large.

As an extra illustration of the performance of \sarge, we present in
\fig{SarFig03} the evaluation of MC-integrals as function of the number of
accepted events. 
Depicted are the integral $\si$ with the bounds on the expected deviation
coming from the estimated expected error, and the relative error. Especially
the graphs with the relative error are illustrative, since they show that
it converges to zero more smoothly for \sarge\ then for \rambo. Notice the
spike for \rambo\ around $\Nac=25000$, where an event obviously hits a
singularity.

\section{Other pole structures}
The APS of (\ref{SarEq001}) is not the only pole structure occurring in the 
squared amplitudes of QCD-processes; not even in purely gluonic processes. 
For example, in the case of $\gl\gl\ra4\gl$, also permutations of 
\begin{equation}
   \frac{1}{\ipb{p_1}{p_3}\ipb{p_2}{p_4}\ipb{\ppin}{p_1}\ipb{\pnip}{p_2}
            \invs{(\ppin-p_1-p_2)}{}}
\label{SarEq017}	    
\end{equation}
occur \cite{Kuijf}. If one is able to generate momenta with this density, it
can be included in the whole density with the use of the adaptive multichannel
technique.
In the interpretation of the transition amplitude as a sum of
Feynman diagrams, this kind of pole structures typically come from $t$-channel
diagrams, which are of the type
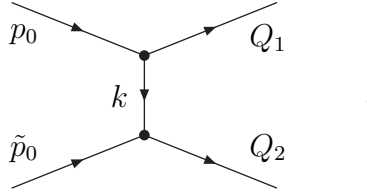
\begin{equation}
\qquad
\parbox{110pt}{\begin{picture}(100,70)(0,0)
\ArrowLine(0,0)(50,20)     \Text(10.0,10.0)[rb]{$\pnip$}
%\LongArrow(12.7,12.7)(29.3,19.3)  
\ArrowLine(50,20)(100,0)   \Text(90.0,10.0)[lb]{$Q_2$}
%\LongArrow(70.7,19.3)(87.3,12.7)  
\Vertex(50,20){2}
\ArrowLine(50,50)(50,20) \Text(44,35)[cr]{$k$}
\Vertex(50,50){2}
\ArrowLine(0,70)(50,50)   \Text(10.0,62.0)[rt]{$\ppin$}
%\LongArrow(14.7,57.3)(31.3,50.7)  
\ArrowLine(50,50)(100,70) \Text(90.0,62.0)[lt]{$Q_1$}
%\LongArrow(70.7,50.7)(87.3,57.3)  
\end{picture}}
\qquad,\notag
\end{equation}
and where, for this case, $Q_1=p_1+p_3$ and $Q_2=p_2+p_4$, so that
$k=p_0-p_1-p_3$. The natural way to generate a density with this pole structure
is by generating $s_i\df\invs{Q}{i}$ with a density proportional to $1/s_i$, a
variable $t$ that plays the role of $\invs{(p_0-p_1-p_3)}{}$, construct with
this and some generated angles the momenta $Q_i$, and then split new momenta
from each of these. 
For $\np=4$, only two momenta have to split off each $Q_i$, and there is a
reasonable simple algorithm to generate these. 

We shall now just present the algorithm, and then show its correctness
using the UAF. If we mention the generation of some random variable $x$ `with a
density $f(x)$' in the following, we mean a density that is proportional to
$f(x)$, and we shall not always write down the normalization explicitly.
Furthermore, $\scm$ denotes the square of the CM-energy and
$\la\df\la(\scm,s_1,s_2)$ the usual Mandelstam variable
\begin{equation}
   \la \;\df\; \scm^2 + s_1^2 + s_2^2 - 2\scm s_1 - 2\scm s_2 - 2s_1s_2  \;\;.
\end{equation}
Of course, a cut has to be implemented in order to generate momenta
following (\ref{SarEq017}), and we shall be able to put $\ipb{p_i}{p_j}>\half
s_0$ for the scalar products occurring in the denominator, where $s_0$ only has
to be larger than zero. To generate the momenta with density (\ref{SarEq017}),
one should
\begin{Alg}[\texttt{T-CHANNEL}]
\begin{enumerate}
\item generate $s_1$ and $s_2$ between $s_0$ and $\scm$ 
      with density $1/s_1$ and $1/s_2$;
\item generate $t$ between $\scm-s_1-s_2\pm\sqrt{\la(\scm,s_1,s_2)}$ 
%     and $\scm-s_1-s_2+\sqrt{\la}$ 
      with density $1/[t(t+2s_1)(t+2s_2)]$;   
\item put $z\lar(\scm-s_1-s_2-t)/\sqrt{\la}$ and 
      generate $\vhi$ uniformly in $[0,2\pi)$;
\item put $Q_1\lar(\sqrt{s_1+\la/(4s)},\sqrt{\la/(4s)}\,\hat{n}(z,\vhi)\,)$ 
      and $Q_2\lar\wcm\enul-Q_1$;
\item for $i=1,2$, generate $z_i>1-4s_0/(t+2s_i)$ with density 
      $1/(1-z_i)$ and $\vhi_i$ uniformly in $[0,2\pi)$, and put
      $q_i\lar\half\sqrt{s_i}\,(1,\hat{n}(z_i,\vhi_i)\,)$; 
\item for $i=1,2$, rotate $q_i$ to the  CMF of $Q_i$, then boost it to the 
      CMF of $Q_1+Q_2$ to obtain $p_i$, and put $p_{i+2}\lar Q_i-p_i$;
\end{enumerate}
\end{Alg}
As a final step, the incoming momenta can be put to
$\ppin\lar\half\wcm(\enul+\ethr)$ and $\pnip\lar\half\wcm(\enul-\ethr)$.
The variables $s_i$ and $z_i$ can easily be obtained by inversion
(cf. \cite{Devroye}). The variable $t$ can best be obtained by generating
$x\df\log(2\sqrt{s_1s_2})-\log t$ with the help of the rejection method
(cf. \cite{Devroye}).
In the UAF, the steps of the algorithm read as follows. Denoting 
\begin{equation}
   \ve_{1}\df\enul+\ethr
   \quad,\qquad
   \ve_{2}\df\enul-\ethr 
   \quad,\qquad
   h_\pm \df \scm-s_1-s_2\pm\sqrt{\la} \;\;,
\end{equation}
and
 \begin{align}
    \textrm{nrm}(\scm,s_1,s_2)
   \;\df&\; \int\frac{dt}{t(t+2s_1)(t+2s_2)}\,
              \theta(h_- < t < h_+) \notag\\
   \;=&\; \frac{1/4}{s_1-s_2}
            \left[  \frac{1}{s_2}\log\frac{1 + 2s_2/h_-}{1 + 2s_2/h_+} 
 	          - \frac{1}{s_1}\log\frac{1 + 2s_1/h_-}{1 + 2s_1/h_+}
	    \right] \;,
\end{align}
we have
\begin{align}
 1.&\;\; \int\frac{ds_1}{s_1}\frac{ds_2}{s_2}\,
         \frac{\theta(s_0<s_{1,2}<\scm)}{(\log\frac{\scm}{s_0})^2} \notag\\
 2.&\;\; \int\frac{dt}{t(t+2s_1)(t+2s_2)}\,
              \frac{\theta(h_- < t < h_+)}
	           {\textrm{nrm}(s,s_1,s_2)} \notag\\
 3.&\;\; \int dz\,\de\left(z-\frac{s-s_1-s_2-t}{\sqrt{\la}}\right)
               \,\frac{d\vhi}{2\pi} \notag\\
 4.&\;\; \int d^4Q_1\,\de\left(Q_1^0-\sqrt{s_1+\sfrac{\la}{4s}}\right)
               \de^3\left(\vec{Q}_1-\sqrt{\sfrac{\la}{4s}}\,
	                                      \hat{n}(z,\vhi)\right)
	       d^4Q_2\,\de^4(Q_1+Q_2-\wcm\enul) \notag\\	
 5.&\;\; \int\prod_{i=1}^2 
          \frac{dz_i}{1-z_i}\,\frac{\theta(1-z_i>\frac{4s_0}{t+2s_i})}
			           {\log\frac{t+2s_i}{2s_0}}\;
          \frac{d\vhi_i}{2\pi}\; 
	  \dfq_i\,\de(q_i^0-\half\sqrt{s_i}\,)\,
                  \de^3(\,\qvec_i-q_i^0\hat{n}(z_i,\vhi_i)\,) \notag \\
 6.&\;\; \int\prod_{i=1}^2 d^4b_i\,\de^4(b_i-\Bo_{Q_i}\ve_{i})\,
                       \de^4(p_i-\Bo_{Q_i}^{-1}\Ro_{b_i}^{-1}q_i)\,
                       \de^4(p_{i+2}+p_i-Q_i) \;\;. \notag
\end{align}
The various assignments imply the following identities. First of all, we have 
\begin{equation}
   \invs{(p_i+p_{i+2})}{} = \invs{Q}{i} = s_i  \;\;.
\end{equation}
Using that $4ss_1+\la=(s+s_1-s_2)^2$ we find
\begin{equation}
   \sqrt{4s}\,(\ip{\ve_1}{Q_1})
   = s+s_1-s_2-z\sqrt{\la} = t+2s_1 
\end{equation}
and the same for $(1\leftrightarrow2)$, so that
\begin{equation}
   t = 4(\ip{\ppin}{Q_1}) - 2\invs{(p_1+p_3)}{}
     = -2\invs{(\ppin-p_1-p_3)}{} \;\;.
\end{equation}
Denote $\Lo_{Q_i}\df\Ro_{b_i}\Bo_{Q_i}$, so that $q_i=\Lo_{Q_i}p_i$. Because
$\Lo_{Q_i}\ve_i\sim\ve_1$, we find that
\begin{equation}
   1-z_i = \frac{2(\ip{\ve_1}{q_i})}{\sqrt{s_i}}
         = 2\frac{(\ip{\ve_1}{\Lo_{Q_i}p_i})}
	         {(\ip{\ve_1}{\Lo_{Q_i}Q_i})}
	 = 2\frac{(\ip{\ve_i}{p_i})}{(\ip{\ve_i}{Q_i})}	\;\;, 
\end{equation}
so that
\begin{equation}
   (t+2s_1)(1-z_1) = 8(\ip{\ppin}{p_1}) 
   \quad\textrm{and}\quad
   (t+2s_2)(1-z_2) = 8(\ip{\pnip}{p_2}) \;\;.
\end{equation}
We can conclude so far that the algorithm generates the correct pole structure.
For the further evaluation of the integrals one can forget about the factors
$s_i$, $t$, $t+2s_i$ and $1-z_i$ in the denominators.
Using that
\begin{equation}
   \dfq_i\,\de(q_i^0-\half\sqrt{s_i}\,)\,
                  \de^3(\,\qvec_i-q_i^0\hat{n}(z_i,\vhi_i)\,)
   \;=\; 2\,\dfq_i\vt(q_i)\,
         \de^3(\frac{2}{\sqrt{s_i}}\,\qvec_i-\hat{n}(z_i,\vhi_i)\,)\;,
\end{equation}
and replacing step 4 by 
\begin{equation}
   \left(\prod_{i=1}^22\sqrt{s_1+\sfrac{\la}{4s}}\,
         d^4Q_i\vt_{s_i}(Q_i)\right)
   \de(z(\vec{Q_1})-z)\,\de(\vhi(\vec{Q_1})-\vhi)\,
   d^4(Q_1+Q_2-\wcm\enul)\;,
\end{equation}
the integrals can easily be performed backwards, i.e., in the order 
$q_i$, $\vhi_i$, $z_i$, $b_i$, $Q_i$, $\vhi$, $z$, $t$, $s_1$, $s_2$. 
The density finally is 
\begin{align}
   \Theta_{\scm}(\{p\}_4)\,
   &\frac{\theta(2\ipb{\ppin}{p_1}>s_0)\,\theta(2\ipb{\pnip}{p_2}>s_0)\,
         \theta(2\ipb{p_1}{p_3}>s_0)\,\theta(2\ipb{p_2}{p_4}>s_0)}
        {\ipb{\ppin}{p_1}\ipb{\pnip}{p_2}\ipb{p_1}{p_3}\ipb{p_2}{p_4}
	 [-\invs{(\ppin-p_1-p_3)}{}]} \notag\\
   &\times 
    \frac{\scm}{24(2\pi)^3}
    \left[\left(\log\frac{s}{s_0}\right)^2
          \log\frac{t+2s_1}{2s_0}\,\log\frac{t+2s_2}{2s_0}\,
	  \textrm{nrm}(\scm,s_1,s_2)\right]^{-1} \;,
\end{align}
where $s_i\df\invs{(p_i+p_{i+2})}{}$ and $t\df-2\invs{(\ppin-p_1-p_3)}{}$.

\section{Appendices}
\subsection*{Appendix A}
We have to calculate the integral
\begin{equation}
   2\scm^2\left(\frac{c^2}{2\pi}\right)^n\int dxd^4b\,
   \de(\invs{b}{}-1)\,\theta(b^0>0)\,\frac{1}{x^{2n+1}}\,
   \exp\left(-\frac{c\wcm}{x}\,b^0\right)
   \;=\;\frac{2\Gamma(2n)\,B(n)}{(2\pi)^n\scm^{n-2}}\;\;, \notag
\end{equation}
where
\begin{equation}
   B(n) \;\df\; \int d^4b\,\de(\invs{b}{}-1)\,\theta(b^0>0)\,(b^0)^{-2n}
        \;=\;   2\pi\int_1^\infty db^0\,(b^0)^{-2n}\,\sqrt{(b^0)^2-1}  
	\;\;.\notag
\end{equation}
The `Euler substitution' $b^0\df\sfrac{1}{2}\,(v^{1/2}+v^{-1/2})$ 
casts the integral in the form
\begin{equation}
   B(n) \;=\; 
   2^{2n-2}\pi\int_1^\infty dv\,\frac{(v-1)^2v^{n-2}}{(v+1)^{2n}}  \;\;.\notag
\end{equation}
By the transformation $v\ra1/v$ it can easily be checked that the integral from
$1$ to $\infty$ is precisely equal to that from $0$ to $1$, so that we may 
write
\begin{equation}
   B(n) \;=\; \frac{2^{2n-2}\pi}{2}\int_0^\infty dv\,
              \frac{v^n-2v^{n-1}+v^{n-2}}{(1+v)^{2n}}
	\;=\; 4^{n-1}\pi\,\frac{\Gamma(n-1)\Gamma(n)}{\Gamma(2n)}\;\;,\notag
\end{equation}
where we have used, by writing $z\df1/(1+v)$, that
\begin{equation}
   \int_0^\infty dv\,v^p(1+v)^{-q} 
   \;=\; \int_0^1 dz\,z^{q-p-2}(1-z)^p 
   \;=\; \frac{\Gamma(q-p-1)\Gamma(p+1)}{\Gamma(q)} \;\;.\notag
\end{equation}

\end{document}